\documentstyle[pra,aps,epsfig,amsmath]{revtex}

\begin{document}
\draft

\twocolumn [ \hsize\textwidth\columnwidth\hsize\csname
@twocolumnfalse\endcsname

\title{Resonant-state solution of the
Faddeev-Merkuriev integral equations for three-body systems with Coulomb
potentials}
\author{Z.\ Papp${}^{1,2}$, J.~Darai${}^{3}$,  C-.Y.\ Hu${}^{1}$,  
  Z.~T.~Hlousek${}^{1}$, B.~K\'onya${}^{2}$ and  
S.\ L.\ Yakovlev${}^{4}$  } 
\address{${}^{1}$ Department of Physics, 
California State University, Long Beach, California 90840 \\
${}^{2}$ Institute of Nuclear Research of the
Hungarian Academy of Sciences, Debrecen, Hungary \\
${}^{3}$ Department of Experimental Physics, University of Debrecen, 
 Debrecen, Hungary \\
${}^{4}$ Department of Mathematical and Computational Physics, 
St.\ Petersburg State University, St.\ Petersburg,  Russia }

\date{\today}

\maketitle

\begin{abstract}
\noindent
A novel method for calculating resonances in three-body
Coulombic systems is proposed. The Faddeev-Merkuriev integral equations
are solved by applying the Coulomb-Sturmian separable expansion method.
The $e^- e^+ e^-$ S-state resonances up to $n=5$ threshold are calculated.
\end{abstract}

\vspace{0.5cm}

]

\section{Introduction}

For three-body systems the Faddeev equations are the fundamental equations.
Three-body bound states correspond to
the solutions of the homogeneous Faddeev equations at real energies,
and resonances, as is usual in quantum mechanics,
are related to complex-energy solutions.

The Faddeev equations were derived for short-range
interactions. However, if we simply plug-in a Coulomb-like potential they become
singular. A formally exact approach was proposed by Noble \cite{noble}. 
His formulation was designed for solving the nuclear three-body
Coulomb problem, where all Coulomb interactions were repulsive.
The interactions were split into short-range and long-range Coulomb-like
parts and the long-range parts were formally included in the 
"free" Green's operator.
Merkuriev extended the idea of Noble by performing the 
splitting in the three-body configuration space \cite{fm-book}.
This was a crucial development since it made possible to treat attractive
Coulomb interactions on an equal footing with repulsive ones.

Recently we have presented a  method  for treating the three-body Coulomb
scattering problem by solving Faddeev-Merkuriev integral equations
using the Coulomb-Sturmian separable expansion technique \cite{phhky}.
We solved the inhomogeneous Faddeev-Merkuriev integral equations for real 
energies. 
Previously, for calculating resonances in three-body systems with short-range 
plus repulsive Coulomb interactions, we solved homogeneous
Faddeev-Noble integral equations by using the Coulomb-Sturmian separable 
expansion technique \cite{zis}. 
In this paper by combining the concepts of Refs.\  \cite{phhky} and \cite{zis}
we solve the homogeneous Faddeev-Merkuriev integral equations for 
complex energies. This way we can
handle all kind of Coulomb-like potentials in resonant-state calculations, 
not only repulsive but also attractive ones. 

In section II we present the homogeneous Faddeev-Merkuriev integral
equations, outlined for systems where two particles out of
the three are identical. 
Many systems, like $e^-e^+e^-$ and $H^-$, fall into this category. 
Then, in section III, we present the 
solution method adapted to the case where all charges have 
the same absolute value. In section 
IV we present our calculations for the $L=0$  resonances
of the $e^-e^+e^-$ system up to the $n=5$ threshold and compare them with the
results of complex scaling calculations \cite{ho}.

\section{Faddeev-Merkuriev integral equations}

The Hamiltonian of a three-body Coulombic  system reads
\begin{equation}
H=H^0 + v_1^C+ v_2^C + v_3^C,
\label{H}
\end{equation}
where $H^0$ is the three-body kinetic energy
operator and $v_\alpha^C$ denotes the
Coulomb-like interaction in the subsystem $\alpha$.
We use throughout the usual configuration-space Jacobi coordinates
$x_\alpha$  and $y_\alpha$. Thus  $v_\alpha^C$ only depends on 
$x_\alpha$ ($v_\alpha^C=v_\alpha^C (x_\alpha)$).
The  Hamiltonian (\ref{H}) is defined in the three-body 
Hilbert space. The two-body potential operators are formally
embedded in the three-body Hilbert space
\begin{equation}
v^C = v^C (x) {\bf 1}_{y},
\label{pot0}
\end{equation}
where ${\bf 1}_{y}$ is a unit operator in the two-body Hilbert space
associated with the $y$ coordinate. We also use the notation 
$X=\{x_\alpha,y_\alpha\}\in {\bf R}^6$.

The role of Coulomb potentials in Hamiltonian (\ref{H}) are twofold.
Their long-distance parts modify the asymptotic
motion, while their short-range parts strongly correlate the
two-body subsystems.
Merkuriev introduced a separation of the three-body
configuration space into different
asymptotic regions. The two-body asymptotic region $\Omega_\alpha$ is
defined as a part of the three-body configuration space where
the conditions
\begin{equation}
|x_\alpha| <  x^0_\alpha ( 1 +|y_\alpha|/ y^0_\alpha)^{1/\nu},
\label{oma}
\end{equation}
with $x^0_\alpha, y^0_\alpha >0$ and $\nu > 2$, are satisfied.
Merkuriev proposed to split the Coulomb interaction  in 
the three-body configuration space into
short-range and long-range terms 
\begin{equation}
v_\alpha^C =v_\alpha^{(s)} +v_\alpha^{(l)} ,
\label{pot}
\end{equation}
where the superscripts
$s$ and $l$ indicates the short- and long-range
attributes, respectively. 
The splitting is carried out with the help of a splitting 
function $\zeta_\alpha$
which possesses the property 
\begin{equation}
\zeta_\alpha(x_\alpha,y_\alpha) \xrightarrow{X_\alpha \to \infty}
\left\{ 
\begin{array}{ll}
1, &  X_\alpha \in \Omega_\alpha \\
0  & \mbox{otherwise.}
\end{array}
\right.
\end{equation}
In practice, in the configuration-space differential equation
approaches, usually the functional form
\begin{equation}
\zeta (x,y) =  2/\left\{1+ 
\exp \left[ {(x/x^0)^\nu}/{(1+y/y^0)} \right] \right\},
\label{oma1}
\end{equation}
was used.

The long-range Hamiltonian is defined as
\begin{equation}
H^{(l)} = H^0 + v_1^{(l)}+ v_2^{(l)}+ v_3^{(l)},
\label{hl}
\end{equation}
and its resolvent operator is
\begin{equation}
G^{(l)}(z)=(z-H^{(l)})^{-1},
\end{equation}
where $z$ is the complex energy-parameter.
Then, the three-body Hamiltonian takes the form
\begin{equation}
H = H^{(l)} + v_1^{(s)}+  v_2^{(s)}+ v_3^{(s)},
\label{hll}
\end{equation}
which formally looks like a three-body Hamiltonian with short-range
potentials. Therefore the Faddeev method is applicable.

In the Faddeev procedure we split the wave function into
three components
\begin{equation}
|\Psi \rangle = |\psi_1 \rangle +
|\psi_2 \rangle +|\psi_3 \rangle,
\end{equation}
where the components are defined by
\begin{equation}
|\psi_\alpha \rangle = G^{(l)} (z) v_\alpha^{(s)} |\Psi \rangle .
\end{equation}
In case of bound and resonant states the wave-function components satisfy 
the homogeneous Faddeev-Merkuriev integral equations
\begin{eqnarray}
 |\psi_1 \rangle &=& G_1^{(l)} (z)
v^{(s)}_1 [ |\psi_2 \rangle + |\psi_3 \rangle ] 
\label{fn-eq1} \\
 |\psi_2 \rangle &=& G_2^{(l)} (z)
v^{(s)}_2 [ |\psi_1 \rangle + |\psi_3 \rangle ] 
\label{fn-eq2} \\
 |\psi_3 \rangle &=& G_3^{(l)} (z)
v^{(s)}_3 [ |\psi_1 \rangle + |\psi_2 \rangle ]
\label{fn-eq3}
\end{eqnarray}
at real and complex energies, respectively.
Here $G^{(l)}_\alpha$ is the resolvent of the channel 
long-ranged Hamiltonian
\begin{equation}
H^{(l)}_\alpha = H^{(l)} + v_\alpha^{(s)},
\label{hla}
\end{equation}
$G^{(l)}_\alpha(z)=(z-H^{(l)}_\alpha)^{-1}$.
Merkuriev has proved that Eqs.\ (\ref{fn-eq1}-\ref{fn-eq3}) 
possess compact kernels,
 and this property remains valid also for
complex energies $z=E-i\Gamma/2$, $\Gamma > 0$.

In atomic three-particle systems the sign of the charge of
two particles are always identical. Let us denote them
by $1$ and $2$, and the non-identical one by $3$.
In this case  $v_3^C$ is a repulsive Coulomb
potential which does not support two-body bound states. Therefore the entire
$v_3^C$ can be considered as long-range potential.
The long-range Hamiltonian is modified as
\begin{equation}
H^{(l)} = H^0 + v_1^{(l)}+ v_2^{(l)}+ v_3^{C}.
\label{hlp}
\end{equation}
Then, the three-body Hamiltonian takes the form
\begin{equation}
H = H^{(l)} + v_1^{(s)}+ v_2^{(s)},
\label{hllp}
\end{equation}
i.e.\ the Hamiltonian of the system appears formally
as a three-body Hamiltonian with two short-range potentials.
Therefore the Faddeev procedure, in this case, gives a 
set of two-component Faddeev-Merkuriev integral equations
\begin{eqnarray}
| \psi_1 \rangle &=  & G_1^{(l)} v_1^{(s)} | \psi_2 \rangle \\
| \psi_2 \rangle &=  & G_2^{(l)} v_2^{(s)} | \psi_1 \rangle.
\label{fm1}
\end{eqnarray}

Further simplification can be achieved if
the particles $1$ and $2$ are identical. Then, the Faddeev components 
$| \psi_1 \rangle$ and $| \psi_2 \rangle$, in their own natural Jacobi
coordinates, have the same functional form
\begin{equation}
\langle x_1 y_1 | \psi_1 \rangle = \langle x_2 y_2 | \psi_2 \rangle
= \langle x y | \psi \rangle.
\end{equation}
Therefore we can  determine $| \psi \rangle$ from the first equation
only 
\begin{equation} \label{fmp}
| \psi \rangle =  G_1^{(l)} v_1^{(s)} p {\mathcal P} 
| \psi \rangle,
\end{equation}
where ${\mathcal P}$ is the operator for the permutation of indexes
$1$ and $2$ and $p=\pm 1$ are eigenvalues of ${\mathcal P}$.
We note that although this integral equation has only one component 
yet gives full account 
on asymptotic and symmetry properties of the system.

\section{Solution method}

We solve these integral equations
by using the Coulomb--Sturmian separable expansion approach \cite{pzwp}.
The Coulomb-Sturmian (CS) functions are defined by
\begin{equation}
\langle r|n l \rangle =\left[ \frac {n!} {(n+2l+1)!} \right]^{1/2}
(2br)^{l+1} \exp(-b r) L_n^{2l+1}(2b r),  \label{basisr}
\end{equation}
with $n$ and $l$ being the radial and
orbital angular momentum quantum numbers, respectively, and $b$ is the size
parameter of the basis.
The CS functions $\{ |n l \rangle \}$
form a biorthonormal
discrete basis in the radial two-body Hilbert space; the biorthogonal
partner defined  by $\langle r |\widetilde{n l}\rangle=
\langle r |{n l}\rangle/r$. 
Since the three-body Hilbert space is a direct product of two-body
Hilbert spaces an appropriate basis
can be defined as the
angular momentum coupled direct product of the two-body bases 
\begin{equation}
| n \nu  l \lambda \rangle_\alpha =
 | n  l \rangle_\alpha \otimes |
\nu \lambda \rangle_\alpha, \ \ \ \ (n,\nu=0,1,2,\ldots),
\label{cs3}
\end{equation}
where $| n  l \rangle_\alpha$ and $|\nu \lambda \rangle_\alpha$ are associated
with the coordinates $x_\alpha$ and $y_\alpha$, respectively.
With this basis the completeness relation
takes the form (with angular momentum summation
implicitly included)
\begin{equation}
{\bf 1} =\lim\limits_{N\to\infty} \sum_{n,\nu=0}^N |
 \widetilde{n \nu l \lambda} \rangle_\alpha \;\mbox{}_\alpha\langle
{n \nu l \lambda} | =
\lim\limits_{N\to\infty} {\bf 1}^{N}_\alpha.
\end{equation}
Note that in the three-body Hilbert space,
three equivalent bases belonging to fragmentation
$1$, $2$ and $3$ are possible.

We make the following approximation on the set of
Faddeev-Merkuriev integral equations
\begin{eqnarray}
 |\psi_1 \rangle &=& G_1^{(l)} (z) {\bf 1}^{N}_1
v^{(s)}_1 [ {\bf 1}^{N}_2 |\psi_2 \rangle +
{\bf 1}^{N}_3 |\psi_3 \rangle ] 
\label{fn-eq1s} \\
 |\psi_2 \rangle &=& G_2^{(l)} (z) {\bf 1}^{N}_2
v^{(s)}_2 [ {\bf 1}^{N}_1 |\psi_1 \rangle +
{\bf 1}^{N}_3 |\psi_3 \rangle ] 
\label{fn-eq2s} \\
 |\psi_3 \rangle &=& G_3^{(l)} (z) {\bf 1}^{N}_3
v^{(s)}_3 [ {\bf 1}^{N}_1 |\psi_1 \rangle +
{\bf 1}^{N}_2 |\psi_2 \rangle ], 
\label{fn-eq3s}
\end{eqnarray}
i.e.\ the short-range potential
$v_\alpha^{(s)}$ in the three-body
Hilbert space is taken to have a separable form, viz.
\begin{eqnarray}
v_\alpha^{(s)} & = &
\lim_{N\to\infty} {\bf 1}^{N}_\alpha v_\alpha^{(s)} {\bf 1}^{N}_\beta
\nonumber \\ 
& \approx & {\bf 1}^{N}_\alpha v_\alpha^{(s)} {\bf 1}^{N}_\beta
= \sum_{n,\nu ,n',
\nu'=0}^N|\widetilde{n\nu l \lambda}\rangle _\alpha \;
\underline{v}_{\alpha \beta }^{(s)}
\;\mbox{}_\beta \langle 
\widetilde{n' \nu' l' \lambda'}|, \label{sepfe}
\end{eqnarray}
where $\underline{v}_{\alpha \beta}^{(s)}=
\mbox{}_\alpha \langle n\nu l \lambda|
v_\alpha^{(s)}|n' \nu ' 
l' \lambda'\rangle_\beta$.
In Eq.~(\ref{sepfe}) the ket and bra states are defined
for different fragmentation, depending on the
environment of the potential operators in the equations.
The validity of this approximation relies on 
the square integrable
property of the terms like $v_\alpha^{(s)} |\psi_\beta \rangle$, which is
guaranteed due to the short range nature of $v_\alpha^{(s)}$.

For solving Eq.\ (\ref{fmp}) we proceed in a similar way,
\begin{equation} \label{fmpa}
| \psi \rangle =  G_1^{(l)}
{\bf 1}^{N}_1 v_1^{(s)} p {\mathcal P} {\bf 1}^{N}_1
| \psi \rangle,
\end{equation}
i.e.\ the operator 
$v_1^{(s)} p {\mathcal P}$ in the three-body
Hilbert space is approximated by a separable form, viz.
\begin{eqnarray}
v_1^{(s)}p {\mathcal P}  & = &
\lim_{N\to\infty} {\bf 1}^{N}_1 v_1^{(s)} p {\mathcal P}  {\bf 1}^{N}_1
\nonumber \\ 
& \approx & {\bf 1}^{N}_1 v_1^{(s)} p {\mathcal P} {\bf 1}^{N}_1 \nonumber \\ 
&  \approx  & \sum_{n,\nu ,n^{\prime },
\nu ^{\prime }=0}^N|\widetilde{n\nu l \lambda}\rangle_1 \;
\underline{v}_1^{(s)}
\;\mbox{}_1 \langle \widetilde{n^{\prime}
\nu ^{\prime} l^{\prime} \lambda^{\prime}}|,  \label{sepfep}
\end{eqnarray}
where $\underline{v}_1^{(s)}=\mbox{}_1 \langle n\nu l \lambda|
v_1^{(s)} p {\mathcal P}  
|n^{\prime }\nu ^{\prime} l^{\prime} \lambda^{\prime}\rangle_1$.
Utilizing the properties of the exchange operator ${\mathcal P}$
these matrix elements can be written in the form $\underline{v}_1^{(s)}= 
p\times \mbox{}_1 \langle n\nu l \lambda| 
v_1^{(s)}|n^{\prime }\nu ^{\prime} l^{\prime} \lambda^{\prime}\rangle_2$.

With this approximation, the solution of Eq.\ (\ref{fmp})
turns into solution of matrix equations for the component vector
$\underline{\psi}_1=
 \mbox{}_1 \langle \widetilde{ n\nu l \lambda} | \psi_1  \rangle$
\begin{equation}
 \{ [ \underline{G}^{(l)}_1(z)]^{-1} - \underline{v}^{(s)}_1 \} 
\underline{\psi}_1 =0,
\end{equation}
where  $\underline{G}_1^{(l)}=\mbox{}_1 \langle \widetilde{
n\nu l\lambda} |G_1^{(l)}|\widetilde{n'\nu' l' \lambda' }\rangle_1$. 
A unique solution exists if
and only if
\begin{equation}
\det \{ [ \underline{G}^{(l)}_1 (z)]^{-1} - \underline{v}^{(s)}_1 \} =0.
\end{equation}

Unfortunately   $\underline{G}_1^{(l)}$  
is not known. It is related to the 
Hamiltonian $H_1^{(l)}$, which itself is a complicated three-body Coulomb
Hamiltonian. In the three-potential formalism \cite{phhky}
$\underline{G}_1^{(l)}$ is linked to simpler quantities via solution of a
Lippmann-Schwinger equation,
\begin{equation}
(\underline{G}^{(l)}_1)^{-1}= 
(\underline{\widetilde{G}}_1)^{-1} -
\underline{U}_1,
\label{gleq}
\end{equation}
where 
\begin{equation}
{\underline{\widetilde{G}}_1}_{ n \nu l \lambda, 
 n^{\prime}\nu^{\prime}l^{\prime} {\lambda}^{\prime}} =
 \mbox{}_1\langle \widetilde{n \nu l \lambda} | 
 \widetilde{G}_1 |
 \widetilde{ n^{\prime}\nu^{\prime}l^{\prime}{\lambda}^{\prime}}
 \rangle_1  
 \label{gtilde}
\end{equation}
and 
\begin{equation}
{\underline{U}_1}_{ n \nu l \lambda,
 n^{\prime}\nu^{\prime} l^{\prime} {\lambda}^{\prime}} =
 \mbox{}_1\langle n\nu l \lambda | U_1 | n^{\prime}\nu^{\prime}
l^{\prime}{\lambda}^{\prime}\rangle_1.
\end{equation}
In our special case, where the sum of the charges of particles $2$ and $3$
is zero, the operator $\widetilde{G}_1$ is the resolvent operator 
of the Hamiltonian
\begin{equation} \label{htilde}
\widetilde{H}_1 = H^{0}+v_1^C,
\end{equation}
and the polarization potential $U_1$ is given by
\begin{equation}
U_1=v_2^{(l)}+v_3^C.
\end{equation}

The most crucial point in this procedure is the
 calculation of the matrix elements
$\underline{\widetilde{G}}_1$, since the  potential
matrix elements $\underline{v}^{(s)}_{1}$ and
$\underline{U}_1$ can always be evaluated numerically by making use of
the transformation of Jacobi coordinates \cite{bb}.
The Green's operator $\widetilde{G}_\alpha$
is a resolvent of the sum of two commuting Hamiltonians,
$\widetilde{H}_1 = h_{x_1}+h_{y_1}$,
where $h_{x_1}=h^0_{x_1}+v_1^C(x_1)$ and
$h_{y_1}=h^0_{y_1}$,
which act in different two-body Hilbert spaces.
Thus, according to the convolution theorem the three-body Green's operator
$\widetilde{G}_\alpha$ equates to
a convolution integral of two-body Green's operators, i.e.
\begin{equation}
\widetilde{G}_1 (z)=
 \frac 1{2\pi {i}}\oint_C
dz^\prime \,g_{x_1}(z-z^\prime)\;g_{y_1}(z^\prime),
 \label{contourint}
\end{equation}
where
$g_{x_1}(z)=(z-h_{x_1})^{-1}$  and
$g_{y_1}(z)=(z-h_{y_1})^{-1}$.
The contour $C$ should be taken  counterclockwise
around the continuous spectrum of $h_{y_1}$
such a way that $g_{x_1}$ is analytic on the domain encircled
by $C$.

To examine the structure of the integrand let us
shift the spectrum of $g_{x_1}$ by
taking  $z=E +{i}\varepsilon$  with
positive $\varepsilon$. By doing so,
the two spectra become well separated and
the spectrum of $g_{y_1}$ can be encircled.
Next the contour $C$ is deformed analytically
in such a way that the upper part descends to the unphysical
Riemann sheet of $g_{y_1}$, while
the lower part of $C$ can be detoured away from the cut
 [see  Fig.~\ref{fig1}]. The contour still
encircles the branch cut singularity of $g_{y_1}$,
but in the  $\varepsilon\to 0$ limit it now
avoids the singularities of $g_{x_1}$.
Moreover, by continuing to negative values of  $\varepsilon$, in order that
we can calculate resonances, the branch cut and pole singularities of
$g_{x_1}$ move
onto the second Riemann sheet of $g_{y_1}$ and, at the same time,
the  branch cut of $g_{y_1}$ moves onto the second Riemann sheet
of $g_{x_1}$. Thus, the mathematical conditions for
the contour integral representation of $\widetilde{G}_1 (z)$ in
Eq.~(\ref{contourint}) can be fulfilled also for complex energies
with negative imaginary part.
In this respect there is only a gradual difference between the
bound- and resonant-state calculations. Now,
the matrix elements $\underline{\widetilde{G}}_\alpha$
can be cast in the form
\begin{equation}
\widetilde{\underline{G}}_1 (z)=
 \frac 1{2\pi {i}}\oint_C dz^\prime \,\underline{g}_{x_1}(z-z^\prime)\;
\underline{g}_{y_1}(z^\prime),
\label{contourint2}
\end{equation}
where the corresponding CS matrix elements of the two-body Green's operators in
the integrand are known analytically for all complex energies (see \cite{phhky}
and references therein),
and thus the convolution integral can be performed also in practice.

\section{Resonant states in positronium ions}

We calculate resonant states
in positronium ion with $L=0$ total angular momentum.
The positronium ion, $\mbox{Ps}^-$ or $e^-e^+e^-$, is a three-body 
Coulomb system that consists of two electrons and one positron. 
We calculate its
resonances by solving Eq.\ (\ref{fmp}). 
We took $x^0=18 \mbox{a}_0$, $y^0=50 \mbox{a}_0$ and
$\nu=2.1$ as the parameters of the splitting function, respectively.

Before presenting our final results we demonstrate the convergence properties 
of this method. In Table (\ref{tab_a}) we show the convergence of 
a resonant state energy with respect to angular momentum channels and
number of Coulomb-Sturmian basis states $N$ in the expansion.
This table shows the accuracy and stability of our calculations.
Table (\ref{tab_b}) contains the final results. For the low-lying resonances
we used CS parameter $b=0.25\mbox{a}_0^{-1}$, and for the high-lying
states we took $b=0.15\mbox{a}_0^{-1}$.
We compare our 
calculation with the result of complex scaling calculations Ref.\ \cite{ho}.
We can report perfect agreements for 
the position of the resonances, but, in most of the cases,
we got much smaller values for the width.

\section{Conclusions}
In this article we have presented a new method for calculating resonances
in three-body Coulombic systems. Our approach is based on the solution of
the homogeneous Faddeev-Merkuriev integral equations for 
complex energies. For this, being an integral equation approach, 
no boundary conditions are needed. We solve the integral equations
by using the Coulomb-Sturmian separable expansion technique. 
The method works equally well for
three-body systems with repulsive and attractive Coulomb interactions.

\acknowledgements
This work has been supported by the NSF Grant No.Phy-0088936
and OTKA Grants under Contracts No.\ T026233 and No.\ T029003. 
We also acknowledge the generous allocation of computer time at the 
San Diego Supercomputing Center by the National Resource Allocation
Committee and at the Department of Aerospace Engineering
of CSULB.  We also greatly appreciate the computing expertise of
the Edinburgh Parallel Computing Centre (EPCC) and  acknowledge the support 
of the European Community Access to Research Infrastructure
action of the Improving Human Potential Programme (contract No
HPRI-1999-CT-00026).

\begin{figure}
\psfig{file=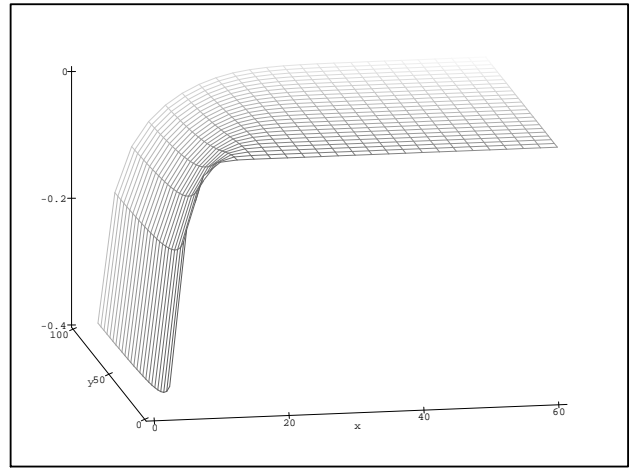,width=7cm,angle=-90}

\caption{Potential $v^{(s)}$, the short-range
part of a $-1/x$ attractive Coulomb potential. }

\label{fig2}
\end{figure}

\begin{figure}
\psfig{file=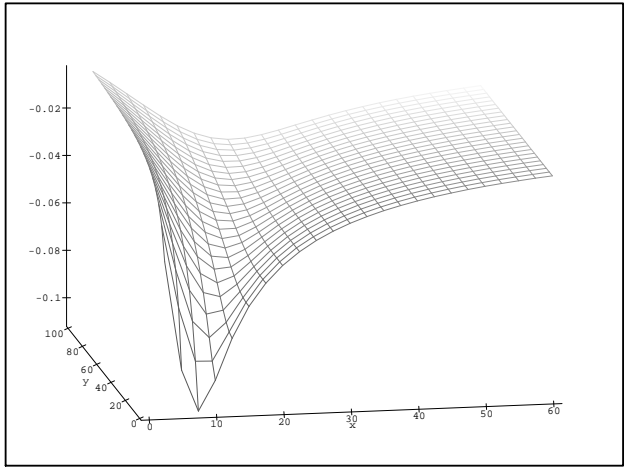,width=7cm,angle=-90}

\caption{Potential $v^{(l)}$, the long-range
part of a $-1/x$ attractive Coulomb potential.}

\label{fig3}
\end{figure}

\begin{figure}
\psfig{file=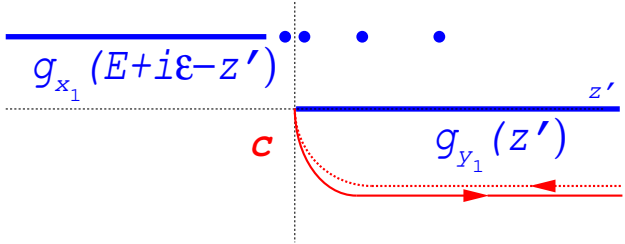,width=8.2cm}

\caption{Analytic structure of $g_{x_1}(z-z^\prime)\;
g_{y_1}(z^\prime)$ as a function of $z^\prime$ with
$z=E+{\mathrm{i}}\varepsilon$, $E>0$, $\varepsilon>0$.
The contour $C$ encircles the continuous spectrum of
$h_{y_1}$. A part of it, which goes on the unphysical
Riemann-sheet of $g_{y_1}$, is drawn by broken line.}
\label{fig1}
\end{figure}

\newpage

\begin{table}
\begin{center}
\begin{tabular}{l|ll}
 $N$  & \multicolumn{1}{c}{$-E_r$}  & \multicolumn{1}{c}{$\Gamma$}   \\ \hline 
 & \multicolumn{2}{c}{$l_{max}=4$}   \\ \hline 
 20  &  0.058667351  & 0.000000133   \\
 21  &  0.058675722  & 0.000000129   \\
 22  &  0.058681080  & 0.000000127   \\
 23  &  0.058684499  & 0.000000127   \\
 24  &  0.058686676  & 0.000000127   \\
 25  &  0.058688060  & 0.000000126   \\ \hline
 & \multicolumn{2}{c}{$l_{max}=5$}   \\ \hline 
 20  &  0.058702010  & 0.000000174   \\
 21  &  0.058710039  & 0.000000170   \\
 22  &  0.058715165  & 0.000000167   \\
 23  &  0.058718426  & 0.000000167   \\
 24  &  0.058720497  & 0.000000167   \\
 25  &  0.058721810  & 0.000000167  \\ \hline
 & \multicolumn{2}{c}{$l_{max}=6$}   \\ \hline 
 20  &  0.058714400  & 0.000000184   \\
 21  &  0.058727373  & 0.000000180   \\
 22  &  0.058727373  & 0.000000177   \\
 23  &  0.058730584  & 0.000000177   \\
 24  &  0.058732621  & 0.000000177   \\
 25  &  0.058733912  & 0.000000177   \\ \hline
 & \multicolumn{2}{c}{$l_{max}=7$}   \\ \hline 
 20  &  0.058717927  & 0.000000188   \\
 21  &  0.058725821  & 0.000000183  \\
 22  &  0.058730852  & 0.000000181   \\
 23  &  0.058734051  & 0.000000180   \\
 24  &  0.058736079  & 0.000000180   \\
 25  &  0.058737364  & 0.000000180   \\ \hline
 & \multicolumn{2}{c}{$l_{max}=8$}   \\ \hline 
 20  &   0.058718914  & 0.000000190    \\
 21  &   0.058726801  & 0.000000186   \\
 22  &   0.058731828  & 0.000000183   \\
 23  &   0.058735023  & 0.000000182  \\
 24  &   0.058737049  & 0.000000183  \\
 25  &   0.058738333  & 0.000000182  \\ \hline
 & \multicolumn{2}{c}{$l_{max}=9$}   \\ \hline 
 20  &   0.058719236  & 0.000000192  \\
 21  &   0.058727121  & 0.000000187  \\
 22  &   0.058732146  & 0.000000185  \\
 23  &   0.058735340  & 0.000000184  \\
 24  &   0.058737366  & 0.000000184  \\
 25  &   0.058738649  & 0.000000184   \\ \hline
 & \multicolumn{2}{c}{$l_{max}=10$}   \\ \hline 
 20  &  0.058719374  & 0.000000193 \\
 21  &  0.058727258  & 0.000000189 \\
 22  &  0.058732283  & 0.000000186 \\
 23  &  0.058735477  & 0.000000185 \\
 24  &  0.058737503  & 0.000000185 \\
 25  &  0.058738786  & 0.000000185
\end{tabular}
\end{center}
\caption{
 Convergence of $^{3}S^{e}$ 3s4s ($L=0$) positronium resonance state, 
 b=0.25.
\label{tab_a}}
\end{table}

\begin{table}
\begin{center}
\begin{tabular}{l|llll}
 State & \multicolumn{2}{c}{Ref.\ \cite{ho}} & \multicolumn{2}{c}{This work}   
 \\ \hline
$^1S^e$ & \multicolumn{1}{c}{$-E_r$}  & \multicolumn{1}{c}{$\Gamma$} 
 & \multicolumn{1}{c}{$-E_r$}  & \multicolumn{1}{c}{$\Gamma$}  \\ \hline       
 2s2s  &  0.1520608 & 0.000086 &  0.1519    & 0.000043   \\
 2s3s  &  0.12730   & 0.00002  &  0.1273    & 0.0000085   \\
 3s3s  &  0.070683  & 0.00015  &  0.0707    & 0.00007   \\
 3s4s  &  0.05969   & 0.00011  &  0.05968   & 0.000053  \\
 4s4s  &  0.04045   & 0.00024  &  0.040428  & 0.00013  \\ 
 4p4p  &  0.0350    & 0.0003   &  0.03502   & 0.00013   \\
 4s5s  &  0.03463   & 0.00034  &  0.03462   & 0.000159   \\
 5s5s  &  0.0258    & 0.00045  &  0.02606   & 0.00010   \\
 5p5p  &  0.02343   & 0.00014  &  0.0234    & 0.00004  \\ \hline
$^3S^e$ & \multicolumn{1}{c}{$-E_r$}  & \multicolumn{1}{c}{$\Gamma$} 
 & \multicolumn{1}{c}{$-E_r$}  & \multicolumn{1}{c}{$\Gamma$}  \\ \hline 
 2s3s &  0.12706    & 0.00001   &  0.127    & 0.000000003   \\
 3s4s  & 0.05873    & 0.00002   &  0.05874  & 0.0000002  \\
 4s5s &  0.03415    & 0.00002   &  0.03420  & 0.0000007   
\end{tabular}
\end{center}
\caption{
Doubly excited $L=0$ resonances of $\mbox{Ps}^-$. The energies and widths are
expressed in Rydbergs.
\label{tab_b}}
\end{table}

{}

\end{document}